\documentclass{article}

\usepackage{arxiv}

\usepackage[utf8]{inputenc} 
\usepackage[T1]{fontenc}    
\usepackage{hyperref}       
\usepackage{url}            
\usepackage{booktabs}       
\usepackage{amsfonts}       
\usepackage{amsmath}
\usepackage{amssymb}
\usepackage{nicefrac}       
\usepackage{microtype}      
\usepackage{graphicx}
\usepackage[numbers,sort&compress]{natbib}
\usepackage{doi}

\title{Cognitive Cells: A Compositional Framework for Populations of Small Language Models}

\author{
	\href{https://orcid.org/0000-0000-0000-0000}
	{\includegraphics[scale=0.06]{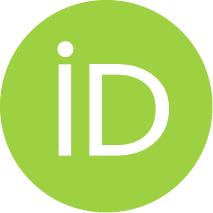}\hspace{1mm}Silvan Ferreira} \\
	Institute of Digital Metropolis \\
	Federal University of Rio Grande do Norte \\
	Natal, RN, Brazil \\
	\texttt{silvan@imd.ufrn.br}
}

\renewcommand{\shorttitle}{Cognitive Cells}

\hypersetup{
pdftitle={Cognitive Cells: A Compositional Framework for Populations of Small Language Models},
pdfsubject={Artificial Intelligence, Cognitive Architectures, Multi-Agent Systems},
pdfauthor={Silvan Ferreira},
pdfkeywords={cognitive architectures, compositional artificial intelligence, multi-agent systems, agentic AI, artificial cognition, cognitive cells},
}

\date{}

\begin{document}
\maketitle

\begin{abstract}
Recent work on large language models and agentic systems raises a basic question that current practice leaves open: how should artificial cognition be decomposed, measured, and composed? We propose studying multi-agent systems from a fixed unit we call a \emph{cognitive cell}: a small, frozen language model with bounded memory and a message interface. The methodological commitment, the \emph{fixed-cell principle}, is to hold this unit constant and vary only the population size, the communication topology, the message bandwidth, and the coordination protocol, so that collective behavior becomes a measurable property of a known device rather than an artifact of per-study engineering. We characterize a single cell by a compact \emph{datasheet} of measurable parameters, and we ask when replicating and connecting cells improves performance: first we measure how one cell behaves alone, then we replicate it and test when voting, communication, and topology help. Instantiating the framework with small frozen models ($1.5$ and $3$ billion parameters), we report a first round of measurements. Adding cells helps only when their errors are not too correlated. A simple correct/incorrect voting model is a useful but conservative null: real open-ended voting can exceed it, because errors are dispersed across many wrong answers rather than concentrated on one. Popular interactive protocols, namely debate, a shared blackboard, and chain revision, do not beat a matched-cost voting baseline in our setting. Finally, a cell's ability to relay several facts, itself a datasheet quantity, predicts whether a population can solve tasks whose evidence exceeds any single cell's memory. We present these as initial measurements within a broader program on scalable artificial cognition, in which multi-agent architectures appear as the special case of cells autonomous enough to be treated as agents.
\end{abstract}


\section{Introduction}
\label{sec:introduction}

The past decade of progress in artificial intelligence has been organized around a single, remarkably productive coordinate system: the neural scaling laws relating model capability to parameters, data, and training compute \cite{kaplan2020scaling, hoffmann2022training}. This coordinate system made the scaling of monolithic models a predictable engineering activity. A second axis of scaling has recently emerged at inference time. It allocates more computation to a frozen model through extended reasoning, sampling, search, and verification \cite{wang2022selfconsistency, snell2024scaling}, and, at the system level, through the coordination of multiple model instances acting as agents \cite{park2023generative, wu2023autogen, hong2023metagpt}. Yet this second axis lacks what the first one has: a shared unit of analysis and predictive laws. Empirical reports on multi-agent systems are contradictory. Some find near-monotonic gains from adding agents \cite{qian2024scaling}; others find non-monotonic returns \cite{chen2024more}, debate that underperforms simple baselines, and consensus that arises from amplified sampling noise \cite{tanaka2026collective}. We argue these findings are mutually consistent: they measure \emph{different devices} under \emph{different conditions}, without a common axis of comparison.

The history of engineering suggests what is missing. Electronics became a compositional discipline once it had a \emph{characterized device}: the transistor, whose datasheet specifies gain, bandwidth, and operating regimes, together with rules for predicting the behavior of arbitrary compositions from that characterization. The datasheet abstracts the underlying physics into a small set of measurable parameters, so that a designer need not reason about semiconductor physics to predict a circuit. Multi-agent artificial intelligence has rich phenomenology and expressive frameworks, but no comparable device characterization. This paper proposes such a device, states candidate laws for how its copies compose, and reports a first round of measurements testing those laws.

A second lineage anticipates this move: cellular automata, which showed that a fixed minimal unit, replicated over a topology under local interaction, suffices to generate the full range of collective behavior, up to universal computation \cite{vonneumann1966theory, gardner1970fantastic, wolfram1984universality}. We raise the abstraction of that unit from a finite-state rule to a bounded cognitive policy; Section~\ref{sec:ca} develops the connection.

This paper proposes the corresponding unit. A \emph{cognitive cell} is a minimal computational unit endowed with explicit interfaces for perception, memory, reasoning, action, goals, and communication, whose internal architecture is \emph{frozen} for the duration of a study. System-level capability is obtained through replication: composing $N$ identical cells over an explicit communication topology under an explicit coordination protocol. The unit stays small; the population carries the scale. The central methodological commitment, which we call the \emph{fixed-cell principle}, is that the cell is a controlled quantity. Only the population size, the topology, the communication budget, and the protocol vary. This commitment is what renders the downstream questions well-posed. The dependence of performance on $N$, the effect of topology, the emergence of specialization, the reliability of collective consensus, and the mapping of tasks onto populations all become measurable properties of a known device. Without the commitment, every such curve would be an artifact of per-study agent engineering.

In plain terms, the program is as follows. We study multi-agent systems starting from a fixed unit, the cognitive cell. We first measure how a cell behaves on its own. We then replicate the cell and test when voting, communication, and topology improve performance. We find that adding cells helps only when their errors are not too correlated; that popular interactive protocols can fail to beat a simple voting baseline at matched cost; and that on tasks requiring distributed memory, communication can create capability that no isolated cell possesses. The remainder of the paper makes this program precise, states it as falsifiable hypotheses, and reports an initial round of measurements with a small frozen model.

In summary, the contributions of this paper are: (i) the cognitive cell, a frozen, versioned, minimal unit of cognition, and the fixed-cell principle that makes composition measurable; (ii) the datasheet, a compact characterization of a cell intended to predict collective behavior; (iii) candidate composition laws for the parallel, serial, and consensual regimes; (iv) an operational test for capability that no single cell possesses, grounded in the voting null model; (v) the task-to-population mapping problem; (vi) seven falsifiable hypotheses with an affordable experimental program; and (vii) an initial round of measurements with a small frozen model that tests the datasheet, the voting null, the value of communication, and the prerequisite for distributed capability. The paper is primarily a framework paper; the measurements of Section~\ref{sec:results} are an initial instantiation, not a complete execution of the program.

Two clarifications delimit the scope. Measurement here serves a design theory, as device characterization serves circuit design; no leaderboard is implied. And the engineering and biological vocabulary makes claims about architecture only, namely the design principle of a repeated frozen unit, with no mechanistic commitment about brains. The claim is narrow: fixing a minimal cognitive unit and studying its laws of composition is a productive way to organize research on scalable artificial cognition, one in which \emph{more is different} \cite{anderson1972more} becomes a quantitative question with measurable order parameters.

\section{Background and Related Work}
\label{sec:related}

The framework draws on four bodies of work: cognitive architectures and language agents, which supply the componential view of a cell; cellular automata, which supply the methodology of a frozen unit replicated over a topology; multi-agent systems of language models, which supply the phenomenology the framework seeks to organize; and the classical theory of voting and reliable computation, which supplies the null models. We review each in turn.

\subsection{Cognitive architectures and language agents}

The decomposition of cognition into interacting functional components has a long history in artificial intelligence, from production systems and classical cognitive architectures such as Soar and ACT-R, synthesized in the standard model of the mind \cite{laird2017standard}, to Minsky's account of intelligence as a society of simple agents \cite{minsky1986society}. Modern language agents rediscovered this structure around a frozen large language model. Chain-of-thought prompting externalized intermediate reasoning \cite{wei2022chain}, ReAct coupled reasoning with tool-mediated action \cite{yao2022react}, Toolformer and successors made tool use a first-class capability \cite{schick2023toolformer}, Reflexion added verbal self-improvement across trials \cite{shinn2023reflexion}, and a substantial literature developed memory mechanisms for persistence across interactions. The authors of \cite{sumers2023cognitive} organize these developments into a cognitive architecture for language agents named CoALA, with modular memories, an action space, and a decision loop. Recent work extends adaptation to the context itself, treating accumulated instructions and strategies as an evolving substrate \cite{zhang2025agentic}. The cognitive cell inherits this componential view and inverts the research question: it asks for the \emph{minimal} unit whose replication scales, and freezes that unit so that composition carries the explanatory burden.

\subsection{From cellular automata to cognitive cells}
\label{sec:ca}

The methodological ancestor of this work is the cellular automaton. Von Neumann asked for the simplest unit which, replicated on a lattice under a fixed local rule, supports globally nontrivial behavior up to self-reproduction \cite{vonneumann1966theory}; the Game of Life exhibited universal computation from a two-state rule \cite{gardner1970fantastic}; and Wolfram turned this into a methodology, holding the unit fixed and classifying the global regimes it produces \cite{wolfram1984universality}. Three commitments define it: the unit is minimal and frozen, interaction is local and explicit, and the object of study is the map from unit, topology, and interaction to global behavior, with phase transitions as its landmarks. Neural cellular automata showed the methodology survives when the local rule becomes a small learned network shared across cells: populations self-organize, regenerate after damage, and act without a central controller, the epistemic stance intact \cite{mordvintsev2020growing}.

The cognitive cell adopts the same methodology with a more expressive unit. The state of a cell is a bounded memory over language and structured data, in place of a symbol from a small finite alphabet. The update rule is a frozen stochastic policy with perception, reasoning, and action, in place of a lookup table. The neighborhood is an arbitrary directed graph with typed, bandwidth-limited message channels, in place of a fixed lattice adjacency. What is preserved from the cellular-automata program is its epistemic stance: a frozen minimal unit, local interaction, scaling by replication, and the study of how unit, topology, and interaction map to collective behavior. Sections~\ref{sec:laws} and~\ref{sec:agenda} carry out this study over populations of cognitive units.

The upgrade also introduces one genuinely new problem that classical cellular automata never faced, and it shapes the entire measurement apparatus of this paper. A Life cell has no individual competence: any global capability of a Life pattern is trivially emergent, because the unit alone can do nothing. A cognitive cell, in contrast, already solves tasks by itself. Collective capability therefore confounds two sources: what the population merely aggregates from individually competent units, and what interaction genuinely creates. Disentangling the two requires an aggregation null model and a criterion for strong emergence, which Section~\ref{sec:capacity} supplies. In this respect the theory of cognitive cells is a theory of cellular automata whose units are strong enough that emergence must be demonstrated instead of assumed.

\subsection{Multi-agent systems of language models}

Frameworks such as AutoGen \cite{wu2023autogen}, MetaGPT \cite{hong2023metagpt}, and generative agent societies \cite{park2023generative} showed that populations of language-model agents can decompose work and exhibit rich social behavior. Quantitative study of \emph{scale} is more recent and mixed: MacNet coordinates over a thousand agents and reports logistic collaborative scaling with an advantage for irregular topologies \cite{qian2024scaling}; sampling-and-voting improves performance broadly \cite{li2024more}, yet returns to additional calls can be non-monotonic \cite{chen2024more}; and debate helps in some regimes \cite{du2023improving} but not against well-tuned single-model baselines in others. Measures of \emph{effective} team size have begun to import the Ringelmann effect, the decline of individual contribution with group size \cite{ringelmann1913recherches}, into these systems. Closest to our program, \cite{tanaka2026collective} builds a minimal statistical-mechanics model separating a drift regime, in which consensus is a lottery, from a selection regime in which weak biases are amplified. We read this literature as converging on the need for a fixed unit and composition laws, which the cognitive cell supplies.

\subsection{Ensembles, juries, and reliable computation from unreliable components}

The oldest quantitative results on collective capability concern voting. The Condorcet jury theorem \cite{decondorcet1785essai} shows that majority vote among independent voters of competence $p > \nicefrac{1}{2}$ approaches certainty as the population grows, while extensions to correlated voters show that positive correlation reduces, and can eliminate, the advantage \cite{ladha1992condorcet}; survey statistics quantifies the same effect through the design effect of cluster sampling \cite{kish1965survey}, and machine learning as ensemble theory. A deeper result is von Neumann's: reliable computation can be synthesized from unreliable components, provided redundancy and majority-based restoring organs are interleaved \cite{vonneumann1956probabilistic}, which licenses capability that aggregation alone cannot deliver, since it concerns computational depth. Populations of cognitive cells sit in both lineages: replicas of one frozen policy are a maximally correlated ensemble, which is why error correlation is a first-class datasheet parameter, and interleaved verification is how populations escape single-trajectory depth limits. The framework departs from ensemble theory in that cells \emph{interact} over rounds, so collective capability becomes a property of a dynamical system in time; quantifying the value of interaction over aggregation is, in our view, the central open measurement problem, for which Section~\ref{sec:capacity} proposes a definition.

\section{The Cognitive Cell}
\label{sec:cell}

We now define the unit. This section gives the formal definition, states the conditions a module must meet to qualify as a cognitive cell, and formulates the fixed-cell principle that turns composition into a measurable question.

\subsection{Definition}

\begin{quote}
\emph{A cognitive cell is a minimal, versioned, frozen unit of cognition: a bounded-memory stochastic policy with explicit interfaces for perception, memory, reasoning, action, goals, and communication.}
\end{quote}

Formally, a cognitive cell is a tuple
\begin{equation}
\mathcal{C} \;=\; \big( \phi,\; \mathcal{M},\; \pi_\theta,\; \mathcal{A},\; \mathcal{G},\; \mathcal{X},\; U \big),
\label{eq:cell}
\end{equation}
whose components realize the six interfaces of the abstract:
\begin{itemize}
	\item \textbf{Perception} $\phi : \mathcal{O} \to \mathcal{E}$ encodes raw observations $o \in \mathcal{O}$, such as text, tool outputs, and environment feedback, into memory-writable events $e \in \mathcal{E}$.
	\item \textbf{Memory} $\mathcal{M}$ is a bounded store with state $h \in \mathcal{H}$, $|h| \le H_{\max}$ tokens, comprising a working context and, optionally, a persistent component surviving across episodes.
	\item \textbf{Reasoning} $\pi_\theta$ is a frozen stochastic policy realized by a small language model with parameters $\theta$, fixed and versioned for the duration of a study, for example \texttt{Cell v0.1}.
	\item \textbf{Action} $\mathcal{A}$ is the space of task-level outputs: final answers, tool invocations, and environment actions.
	\item \textbf{Goals} $\mathcal{G}$ is the space of declarative goal specifications $g$, loaded into memory at episode start and treated as data; the architecture of the cell is unaffected by the goal.
	\item \textbf{Communication} $\mathcal{X} = (\mathbb{M}, \beta)$ specifies a typed, schema-constrained message space $\mathbb{M}$ and a per-round bandwidth constraint $|m| \le \beta$ tokens.
	\item \textbf{Update} $U : \mathcal{H} \times \mathcal{E}^{*} \times \mathbb{M}^{*} \times \mathcal{A} \to \mathcal{H}$ governs how perceived events, incoming messages, and the cell's own outputs are consolidated into the next memory state, respecting $H_{\max}$.
\end{itemize}

A single cell operating in isolation reduces to a familiar language agent: goal in memory, a reason-and-act loop, bounded context. The abstraction earns its name at $N > 1$.

\subsection{What makes a cell cognitive}
\label{sec:normative}

The tuple of Equation~\ref{eq:cell} is permissive by itself, and a definition that admitted any module with input, state, and output would drain the term of content. We therefore state the qualification normatively. A unit qualifies as a cognitive cell only if it satisfies four conditions jointly. First, \emph{goal conditioning}: its behavior is modulated by a goal specification $g$ loaded as data, so the same unit pursues different objectives under different goals. Second, \emph{bounded persistent state}: it maintains a memory within $H_{\max}$ whose contents influence future behavior, so the unit has a history and can be individuated by it. Third, \emph{a stochastic policy over actions and messages}: it selects among actions and emits communication under uncertainty, so behavior is a draw from a distribution and diversity is measurable. Fourth, \emph{a bounded, versioned interface}: its interaction with the world passes exclusively through the declared perception, action, and message channels, so the unit is replicable under a fixed published specification and its datasheet is well-defined.

These conditions exclude a stateless function (no persistent state to individuate it), a fixed-rule router (no goal conditioning and no policy), and a module coupled to peers through unbounded activation exchange (no sealed interface, so replication and characterization lose meaning). They admit a bounded verifier, a single-step proposer, or a learned router, provided each carries goals, state, a policy, and a sealed interface. This is how a cell can be far smaller than an agent while remaining a unit of cognition.

\subsection{The fixed-cell principle}

The defining methodological commitment of the framework is that the tuple of Equation~\ref{eq:cell} is \emph{frozen}: published, versioned, and held constant across an entire experimental campaign. The free variables of a study are the population size $N$, the communication topology $G$, the bandwidth $\beta$, the coordination protocol $P$, and the task distribution $\mathcal{T}$. This is a deliberate inversion of common practice in multi-agent research, where the unit itself, including prompts, roles, memories, and orchestration, is redesigned per paper. The consequence of that practice is that results do not compose across studies and the field accumulates phenomenology in place of laws. Under the fixed-cell principle, every measured curve is a property of a known device. The cost of the principle is breadth: no per-task agent engineering is permitted. That cost is the point. It converts claims of the form ``our system beats baselines'', which are unfalsifiable across unit redesigns, into claims of the form ``this device obeys this law'', which are falsifiable.

Multi-agent architectures are the special case in which cells are autonomous enough, with rich goals, persistent memory, and broad action spaces, to be treated as agents. The generalization is downward: a cell may be far more modest, such as a bounded verifier, a single-step proposer, or a router, and autonomy becomes a designed property of the unit rather than a definitional prerequisite. What this buys is what transistors bought circuit design: the freedom to ask how much capability resides in the \emph{organization}, in $N$, $G$, $\beta$, and $P$, with the unit held constant, and the ability to answer because the unit does not move.

\section{Composition}
\label{sec:composition}

\subsection{Populations as stochastic dynamical systems}

A \emph{population} is a directed graph $G = (V, E)$ with $|V| = N$ cells. At interaction round $t$, cell $i$ holds memory state $h_i^t$, perceives $u_i^t = \phi(o_i^t)$, receives messages $\{ m_{j \to i}^t : (j,i) \in E \}$ from its in-neighbors, and emits actions and messages
\begin{equation}
\big( a_i^t,\; \{ m_{i \to j}^t \}_{(i,j) \in E} \big) \;\sim\; \pi_\theta\!\left( \cdot \,\middle|\, h_i^t \right),
\qquad
h_i^{t+1} \;=\; U\!\left( h_i^t,\; u_i^t,\; \{ m_{j \to i}^t \},\; a_i^t \right).
\label{eq:dynamics}
\end{equation}
The population is therefore a stochastic dynamical system on the product space $\mathcal{H}^N$, parameterized by $(\theta, N, G, \beta, P, T_{\mathrm{rounds}})$. All cells share $\theta$. Heterogeneity, if it arises, arises in the memory states $h_i$, a fact that Section~\ref{sec:laws} elevates into a hypothesis about the origin of specialization.

Given a task $T \sim \mathcal{T}$ with success criterion $R_T$, define collective performance and total cost as
\begin{equation}
\Psi(N, G, P; \mathcal{T}) \;=\; \mathbb{E}_{T, \text{rollouts}}\!\left[ R_T \right],
\qquad
K(N, G, P) \;=\; N \, c_{\mathrm{inf}} \, T_{\mathrm{rounds}} \;+\; |E| \, c_{\mathrm{msg}} \, T_{\mathrm{rounds}},
\label{eq:psi-cost}
\end{equation}
where $c_{\mathrm{inf}}$ is the per-round inference cost of one cell and $c_{\mathrm{msg}}$ the marginal cost of emitting and ingesting a message. The empirical objects of the entire program are the surfaces $\Psi$ and $\nicefrac{\Psi}{K}$ over the design space $(N, G, \beta, P)$, organized by task family.

\subsection{Protocols}

A coordination protocol $P$ specifies scheduling, whether synchronous in rounds or asynchronous by events; role assignment if any; the aggregation of cell outputs into a system-level answer, whether by voting, by selection through a designated cell, or by pipeline hand-off; and termination. Three protocol archetypes anchor the analysis, corresponding to the three regimes of Section~\ref{sec:laws}. In \emph{parallel} protocols, cells attack sub-problems or independent samples and an aggregator combines results. In \emph{serial} protocols, cells form processing stages and each consumes the output of the previous. In \emph{consensual} protocols, cells exchange beliefs over rounds until the population commits to a collective answer. Realistic systems mix the archetypes. The archetypes are studied in pure form because their laws differ in functional form, with the aim that laws for mixed protocols be derivable from those for the pure ones.

A scope note is in order. Throughout this paper, protocols are \emph{exogenous}: the runtime imposes $P$ on the population, in the way a test bench imposes stimuli on a circuit, and the population is measured under the imposed protocol. Questions of protocol origin, such as who decomposes a task, who assigns work at large $N$, and how coordination could be generated by the cells themselves during reasoning, are deliberately excluded here, both because a device theory requires controlled operating conditions and because endogenous coordination deserves, and will receive, a dedicated treatment in subsequent work.

\section{Quantifying Cognitive Capacity}
\label{sec:capacity}

We quantify capacity at two levels. At the individual level, a cell is summarized by a datasheet of measurable parameters. At the collective level, we ask what replication alone achieves, using a null model for voting, and then define the value of interaction and a test for capability that only interaction can create.

\subsection{Individual capacity: the datasheet}

The datasheet\footnote{The term is used in the electronic-engineering sense of a component specification sheet. It is unrelated to datasheets for datasets \cite{gebru2021datasheets}, which document the provenance and composition of training corpora.} of a cell is a low-dimensional vector of measurable parameters, estimated from cheap pilot runs on a task family $\mathcal{T}$, intended to predict collective behavior without reference to the cell's internals:
\begin{equation}
d(\mathcal{C}; \mathcal{T}) \;=\; \big( p,\; D,\; \rho,\; \varphi,\; \eta,\; w \big),
\label{eq:datasheet}
\end{equation}
whose components are defined, measured, and interpreted in Table~\ref{tab:datasheet-def}. Two points matter throughout. First, error correlation $\rho$ is the load-bearing parameter: because all cells share $\theta$, $\rho > 0$ generically, and it is the main obstruction to naive collective gains. Second, a cell has no single universal datasheet; it has a \emph{task-family-indexed} datasheet $d(\mathcal{C}; \mathcal{T})$, reported per family and studied for its stability across families.

\begin{table}[t]
\centering
\small
\caption{The datasheet parameters: definition, measurement, and what each predicts. The datasheet is indexed by task family, $d(\mathcal{C};\mathcal{T})$, not a single universal vector.}
\label{tab:datasheet-def}
\begin{tabular}{@{}clp{3.9cm}p{3.1cm}p{2.9cm}@{}}
\toprule
Sym. & Name & Definition & Measurement & Predicts \\
\midrule
$p$ & competence & single-cell success probability on $\mathcal{T}$ & accuracy over instances & baseline accuracy \\
$D$ & diversity & entropy of the answer distribution under resampling & entropy over resamples of one instance & headroom for voting gains \\
$\rho$ & error correlation & pairwise correlation of correctness across replicas & repeated sampling on the same instances & voting saturation ($N\!\approx\!1/\rho$) \\
$\varphi$ & comm.\ fidelity & rate a received fact is correctly used & fact-injection probes & value of communication \\
$\eta$ & persuadability & shift in answers under a peer's advocacy & advocacy probes & consensus drift vs.\ selection \\
$w$ & memory span & context load beyond which $p$ degrades & distractor injection & distributed-memory limits \\
\bottomrule
\end{tabular}
\end{table}

The scientific bet of the framework, stated as Hypothesis H2 below and treated as falsifiable, is that this six-parameter datasheet is approximately \emph{sufficient}: that $\Psi(N, G, P)$ is predictable from $d$ and coarse task structure, with residuals that localize in identifiable interaction effects. Where the bet fails, the failure is itself a result. It identifies the dimensions along which low-dimensional device characterization is provably inadequate for cognitive units, in contrast to electronic ones.

\subsection{Individuation: how identical cells differ}
\label{sec:individuation}

An immediate objection deserves a direct answer: if every cell shares the same parameters $\theta$, in what sense does a population contain more than one of them? The answer is that $\theta$ fixes a distribution over behaviors, and individuation arises from three sources that the parameters do not determine. First, sampling: $\pi_\theta$ is stochastic, and each cell is an independent draw of trajectories from it, so two cells given the same input explore different reasoning paths, commit different slips, and reach different conclusions whenever the decoding temperature is positive. Second, context: cells in a population occupy different positions in $G$, receive different sub-problems, and read different messages, and identical parameters conditioned on different inputs produce different behavior. Third, history: the memory states $h_i$ evolve under each cell's private trajectory, and initially identical memories diverge over rounds, a divergence that Hypothesis H4 promotes to the candidate mechanism of specialization. The shared parameters fix a distribution of behavior; the divergent memory states are what make cells distinct within it.

This has a quantitative counterpart. A cell's error on an instance splits into a \emph{systematic} component, shared by every replica of $\theta$, and a \emph{stochastic} component, private to the trajectory. Replication attacks only the latter: averaging cancels sampling noise but leaves the shared bias, as averaging $N$ voltmeters of one factory model cancels reading noise but preserves the calibration error. The datasheet parameter $\rho$ is the shared fraction of error variance, and the effective-population law below is its accounting. At temperature zero with identical inputs, $\rho = 1$ and the population collapses to a single effective cell; as sampling and contextual differentiation grow, $\rho$ falls and effective size rises toward $N$.

\subsection{Collective capacity and the effective-population law}

The first composition law follows from $\rho$ alone. For majority aggregation of $N$ exchangeable, positively correlated voters, the design effect of cluster sampling \cite{kish1965survey} gives an \emph{effective population size}
\begin{equation}
N_{\mathrm{eff}} \;=\; \frac{N}{1 + (N - 1)\,\rho},
\label{eq:neff}
\end{equation}
so that collective accuracy behaves approximately as a Condorcet jury of $N_{\mathrm{eff}}$ \emph{independent} voters of competence $p$ \cite{decondorcet1785essai, ladha1992condorcet}. By the normal approximation to the binomial, the entire curve of accuracy against $N$ for aggregation-style protocols collapses onto a two-parameter predictor:
\begin{equation}
P_{\mathrm{maj}}(N) \;\approx\; \Phi\!\left( \sqrt{N_{\mathrm{eff}}}\; \frac{2p - 1}{2\sqrt{p(1-p)}} \right),
\label{eq:jury}
\end{equation}
with $\Phi$ the standard normal distribution function. We stress the scope of Equation~\ref{eq:jury}. It is the \emph{simplest null model} for aggregation, and it applies when only the binary distinction correct/incorrect matters, so that a wrong collective decision requires the erring cells to agree on the \emph{same} wrong answer. On open-ended tasks, where errors are dispersed across many alternatives, the plurality of correct answers can win even when a minority of cells is correct; the appropriate null must then account for the \emph{dispersion} of errors, governed by the diversity $D$, and not only for their rate. We therefore treat Equation~\ref{eq:jury} as a conservative lower bound rather than as ``the'' aggregation law, and Section~\ref{sec:results} reports measurements in which real voting exceeds it for exactly this reason. Treated as such a bound, Equation~\ref{eq:neff} still makes an immediate qualitative point: gains saturate at $N \approx \nicefrac{1}{\rho}$, since $N_{\mathrm{eff}} \to \nicefrac{1}{\rho}$ as $N \to \infty$. The apparent contradiction in the empirical literature thereby becomes mutually consistent under a single measurable parameter. The statements ``$N{=}10$ helped'' and ``$N{=}10$ did not help'' describe devices with different $\rho$. This consistency is a property of the model; whether real systems carry the values of $\rho$ that the reading requires is an empirical question, and the experimental program is designed to answer it. A falsifiable prediction follows: interventions that reduce $\rho$, such as sampling temperature, prompt perturbation, heterogeneous memory states, and disagreement-seeking protocols, must shift the saturation point rightward by a quantitatively predictable amount. We regard the experimental verification of Equations~\ref{eq:neff} and~\ref{eq:jury} on frozen cells, together with the measurement of the response of $\rho$ to such interventions, as the natural first anchor result of the program.

\subsection{The value of interaction, and a test for emergence}

Equation~\ref{eq:jury} is a \emph{null model}: it describes what replication plus aggregation achieves without any communication. The value of interaction is then operationally defined as the excess of observed collective performance over the null model at matched cost,
\begin{equation}
\Delta\Psi_{\mathrm{comm}}(N, G, P) \;=\; \Psi_{\mathrm{observed}}(N, G, P) \;-\; P_{\mathrm{maj}}\!\left(N;\, p, \rho\right),
\label{eq:deltapsi}
\end{equation}
computed against the resampling-and-voting baseline of a single cell. A positive $\Delta\Psi_{\mathrm{comm}}$ certifies that message exchange adds information beyond independent evidence accumulation. A negative $\Delta\Psi_{\mathrm{comm}}$ certifies that interaction destroys it, whether through error propagation, conformity, or drift. Much of the disagreement in the multi-agent literature dissolves, we contend, once results are reported as $\Delta\Psi_{\mathrm{comm}}$ in place of raw wins over ad hoc baselines. The quantity also supplies the framework's answer to the reasonable objection that a single large model may outperform any population of small cells. The comparison that matters is conducted on the cost frontier of Section~\ref{sec:laws}, and the question of \emph{where} in task space each regime dominates is a primary object of study for the framework.

\label{sec:emergence}
The null model also yields a test for the strongest claim a compositional theory can make: that interaction \emph{creates} capability absent from every constituent. Consider task families constructed so that single-cell competence is structurally null, $p \approx 0$, for architectural reasons independent of skill or effort. Two constructions guarantee this. In the first, the state required to represent the problem exceeds the memory bound: the answer demands the joint consideration of $k$ evidence fragments with $k \cdot |\text{fragment}| \gg H_{\max}$, so no single cell can hold the relevant state, whatever its skill. In the second, the required inferential depth exceeds single-trajectory reliability: a dependency chain of length $L$ with per-step error $\varepsilon$ succeeds end-to-end with probability $(1-\varepsilon)^L$, which is driven arbitrarily close to zero by construction.

In this regime the null model inherits the nullity of its inputs: Equation~\ref{eq:jury} with $p \approx 0$ yields $P_{\mathrm{maj}} \approx 0$ for every $N$, since aggregation concentrates existing competence and cannot manufacture it. Consequently, any measured $\Psi > 0$ is attributable \emph{in its entirety} to interaction, and $\Delta\Psi_{\mathrm{comm}}$ ceases to be an excess and becomes a decisive test:
\begin{equation}
p \approx 0 \;\;\wedge\;\; \Psi(N, G, P) \gg 0 \quad \Longrightarrow \quad \text{strong emergence, relative to the aggregation null.}
\label{eq:emergence}
\end{equation}
Several mechanisms could produce capability in this regime: distributed working memory, in which cells jointly hold $N \cdot H_{\max}$ of state that none can hold alone and exchange the evidence fragments; interleaved restoration, in which verifier cells clean errors before they propagate \cite{vonneumann1956probabilistic}; and a generate-and-verify asymmetry, in which weak proposers coupled to a strong verifier accumulate capability over rounds, with a precedent in mathematical discovery by program search over model populations \cite{romeraparedes2024mathematical}. The test of Equation~\ref{eq:emergence} makes emergence measurable rather than rhetorical, and Section~\ref{sec:agenda} builds task families to probe it.

\section{Composition Laws and Operating Regimes}
\label{sec:laws}

The three protocol archetypes give three composition laws, each with a characteristic failure mode: saturation under error correlation in the parallel regime, error accumulation in the serial regime, and unreliable consensus in the consensual regime. We state each in turn, then discuss specialization and the cost frontier against a single large model.

\subsection{Parallel regime: logistic growth and decomposability}

For decomposable tasks, those solvable as parallel sub-solutions plus cheap verification or aggregation, collective performance is expected to follow logistic growth in population size,
\begin{equation}
\Psi(N) \;=\; \frac{\Psi_{\max}}{1 + e^{-k (\ln N - \ln N_0)}},
\label{eq:logistic}
\end{equation}
consistent with the collaborative scaling reported at the thousand-agent scale \cite{qian2024scaling}. The framework's added claim is that the midpoint $N_0$ and steepness $k$ are functions of the datasheet and of the task's decomposition width, that is, the number of independently attackable sub-problems, in place of free phenomenological constants. Saturation is governed by Equation~\ref{eq:neff} when sub-solutions are aggregated by vote, and by verification fidelity $\varphi$ when they are aggregated by checking.

\subsection{Serial regime: cascades, error accumulation, and restoration}

Serial protocols place cells in a processing chain of length $L$, where each stage consumes the previous stage's output. Errors then compound multiplicatively. With per-stage error $\varepsilon_l$,
\begin{equation}
\varepsilon_{\mathrm{chain}}(L) \;=\; 1 - \prod_{l=1}^{L} (1 - \varepsilon_l) \;\approx\; 1 - (1 - \varepsilon)^L,
\label{eq:cascade}
\end{equation}
the familiar multiplicative accumulation of errors along a cascade of stages \cite{friis1944noise}, whose lesson transfers intact: \emph{early-stage} quality dominates the composite result. The design rule that follows is to allocate verification capacity preferentially to early stages, or to interleave verifier cells whose competence exceeds the proposers' at the point of insertion.

Von Neumann's synthesis of reliable organisms from unreliable components \cite{vonneumann1956probabilistic} sharpens the rule into a quantitative construction. Interleave a \emph{restoring stage} every $k$ steps: $r$ verifier cells check the partial result and a local majority corrects it before propagation resumes. Provided the restoring stage's own error is below the accumulated segment error it corrects, end-to-end reliability can be held bounded for arbitrary chain length $L$, at a multiplicative cost overhead of order $\nicefrac{r}{k}$. The construction yields a testable prediction with no free parameters beyond the datasheet: the optimal restoration spacing $k^{*}$ is computable from the per-step error $\varepsilon$ and the restorer's fidelity, and populations instrumented at $k^{*}$ should reach inferential depths at which every single-trajectory baseline fails with near certainty. This is the serial-regime instantiation of the strong-emergence criterion of Section~\ref{sec:emergence}.

\subsection{Consensual regime: drift, selection, and the reliability of collective conclusions}

Consensual protocols raise the sharpest scientific question of the program: when a population of cells converges on an answer, is the convergence an \emph{inference} or a \emph{lottery}? Model each cell's belief as a point $x_i \in \Delta^k$ on the probability simplex over $k$ options, updated by gossip. Upon observing a sampled output $a_j$ of a neighbor, the cell updates $x_i \leftarrow (1 - \eta)\, x_i + \eta\, e_{a_j}$, with $\eta$ the persuadability of the datasheet and $e_{a}$ the vertex of the chosen option. Because each cell learns from \emph{samples} of its neighbors' beliefs, one cell's arbitrary choice becomes the next cell's evidence, and sampling noise of order $\sqrt{\nicefrac{\eta^2}{N}}$ per interaction can compound into full consensus even when no cell holds any intrinsic preference. This is the memetic-drift regime identified in \cite{tanaka2026collective}, by analogy with neutral evolution. When intrinsic biases or task evidence are sufficiently strong relative to the sampling noise, the system instead operates in a selection regime, in which weak signals are amplified and the correct option reliably wins. The control parameters are the population size $N$, the bandwidth $\beta$, the persuadability $\eta$, and the cells' intrinsic uncertainty, which plays the role of a temperature. The order parameter is the consensus magnitude $\lVert \bar{x} \rVert$. Mapping the drift--selection phase boundary for frozen cells, which is cheap to do with naming games and choice tasks at small scale, is one of the program's flagship experiments. Its practical stake is plain: deployed collectives whose operating point lies on the drift side of the boundary produce confident collective conclusions that are, in a precise sense, noise.

\subsection{Specialization as symmetry breaking}

Cells are identical in $\theta$ and individuated by memory. The population dynamics of Equation~\ref{eq:dynamics} act on the joint memory state $(h_1, \dots, h_N)$, and nothing forces the memory states to remain exchangeable. A central hypothesis of the framework is that stable division of labor can emerge from memory divergence alone, as an order--disorder transition in which role structure appears above a critical combination of task pressure and population size, measurable by the entropy of the empirical role distribution. A positive result would establish that heterogeneous \emph{experience} suffices for specialization in cognitive collectives, and that model heterogeneity is dispensable, connecting the framework to the literature on agent memory as the substrate of individuality. A negative result would locate the missing ingredient. Either outcome disciplines the currently ad hoc practice of hand-assigning roles.

\subsection{The cost frontier and the crossover with monolithic scale}

The honest comparison against a single large model is conducted on the frontier
\begin{equation}
\Psi^{*}(B) \;=\; \max_{N, G, P} \;\Psi(N, G, P) \quad \text{s.t.} \quad K(N, G, P) \le B,
\label{eq:frontier}
\end{equation}
swept over budgets $B$ and plotted against monolithic models evaluated at matched inference cost. The regime analysis above predicts a partition of task space. Populations of frozen small cells should dominate on tasks with large decomposition width, on tasks rewarding diverse hypothesis generation, and in the strong-emergence regime of Section~\ref{sec:emergence}, where required state exceeds any single context. Monoliths should dominate on tasks requiring long serial dependency chains within a single coherent context and without checkable intermediate states, where Equation~\ref{eq:cascade} punishes the population and restoration has nothing to verify against. Charting this crossover boundary, parameterized at first approximation by decomposition width and serial depth, is a primary deliverable of the program, and one well suited to resource-constrained academic groups, since the question of how much capability $N$ frozen small cells can reach is neglected by laboratories whose economics favor monolithic scale.

\section{The Compilation Problem}
\label{sec:compilation}

Once the datasheet and the composition laws exist, system construction becomes an optimization. Given a task $T$ with estimated structure, namely decomposition width, serial depth, and verification cost, a cell with datasheet $d$, and a budget $B$, choose
\begin{equation}
(N^{*}, G^{*}, P^{*}) \;=\; \arg\max_{N, G, P} \;\widehat{\Psi}(N, G, P;\, d, T)
\quad \text{s.t.} \quad K(N, G, P) \le B,
\label{eq:compilation}
\end{equation}
where $\widehat{\Psi}$ is the predictive model assembled from Equations~\ref{eq:neff} to~\ref{eq:cascade} and their empirical refinements. We call this the \emph{compilation problem}: the mapping of tasks to circuits of cells, in strict analogy with hardware compilation. Its deliverables are twofold. The theoretical deliverable asks how accurate $\widehat{\Psi}$ can be from datasheet-level information alone, which is a precise restatement of Hypothesis H2. The practical deliverable is an open-source runtime that spawns, connects, and schedules cell populations, together with the compiler layer above it. The runtime is the program's infrastructural investment. Every experiment described in this paper runs on it, and it is the artifact through which the framework becomes adoptable and the reported laws reproducible by other groups.

\section{Learning to Compose}
\label{sec:learning}

The program so far treats $\theta$ as frozen. A natural extension, which we flag here as future work rather than develop, trains one shared $\theta$ against a \emph{collective} reward $J(\theta) = \mathbb{E}_{G, T, \text{rollouts}}[R_{\mathrm{collective}}]$, optimizing the cell to be a good \emph{colleague} rather than a good soloist. Verifiable-reward reinforcement learning \cite{shao2024deepseekmath, deepseekai2025deepseekr1} and counterfactual, marginal-contribution credit assignment \cite{shapley1953value} are natural tools. The questions of interest are whether collective training lowers the error correlation $\rho$, shifting the saturation point of Equation~\ref{eq:neff} rightward so that the population \emph{learns diversity}; whether it raises communication fidelity $\varphi$ under the bandwidth constraint $\beta$; and whether a cell trained in small populations generalizes to larger $N$. Parameter sharing keeps this within modest compute, since one small model is trained and populations are inference-time replicas.

\section{Research Agenda: Hypotheses and Experimental Program}
\label{sec:agenda}

The framework's claims are stated as seven falsifiable hypotheses, followed by an experimental program that tests them with modest compute; Section~\ref{sec:results} reports initial results from that program.

\subsection{Falsifiable hypotheses}

The hypotheses divide into two classes. Composition-law hypotheses, namely H1, H2, H3, and H5, assert that collective behavior is predictable from device parameters; their failure would show that cognitive units resist low-dimensional characterization. Emergence hypotheses, namely H4, H6, and H7, assert that composition creates properties absent from the unit, including specialization, learned diversity, and capability itself; their confirmation would establish that the population is a cognitive object in its own right, with the aggregation null of Section~\ref{sec:emergence} as the standard of proof throughout.

\begin{description}
	\item[H1. Composability threshold and regime split.] For each task family there exists a competence--diversity threshold. Above it, aggregation gains are bounded below by the correct/incorrect voting model of Equations~\ref{eq:neff} and~\ref{eq:jury} and saturate near $N \approx \nicefrac{1}{\rho}$, with real open-ended voting exceeding the bound by an amount that grows with the diversity $D$. Below the threshold, and in particular at $p \approx 0$, no population size yields gains under aggregation alone; whatever capability appears there is governed by H7.
	\item[H2. Datasheet sufficiency.] The six-parameter datasheet of Equation~\ref{eq:datasheet} predicts $\Psi(N, G, P)$ within stated error for a fixed task family, with residuals localizing in identifiable interaction effects; where it fails, the failure dimensions are stable and nameable.
	\item[H3. Regime structure.] Task decomposability determines the functional form of $\Psi(N)$: logistic as in Equation~\ref{eq:logistic} for parallel protocols, cascade decay as in Equation~\ref{eq:cascade} for serial protocols, and phase-transition-like for consensual protocols, with parameters computable from the datasheet.
	\item[H4. Specialization from memory alone.] Populations of cells identical in $\theta$ and divergent in memory develop stable role structure above a critical boundary in the plane of task pressure and population size. Emergent role structure raises collective performance above that of any homogeneous population of the same size, and the role repertoire held simultaneously by the population exceeds the repertoire expressible by any single cell.
	\item[H5. Cost crossover.] There is a nontrivial region of task space in which a population of frozen small cells strictly dominates a single large model at matched inference budget, in the sense of Equation~\ref{eq:frontier}; its boundary is predictable from decomposition width and serial depth.
	\item[H6. Collective training.] Reinforcement learning with collective reward on a shared-parameter cell (Section~\ref{sec:learning}) reduces $\rho$ and raises $\Delta\Psi_{\mathrm{comm}}$, with transfer across population sizes and topologies. The population, in this sense, learns the diversity that emergence requires.
	\item[H7. Strong emergence.] There exist task families, constructed so that single-cell competence is structurally null because required state exceeds $H_{\max}$ or required inferential depth exceeds single-trajectory reliability, in which interacting populations achieve $\Psi$ substantially above zero. Since the aggregation null of Equation~\ref{eq:jury} is itself null in this regime, the entirety of measured capability is attributable to interaction, satisfying the test of Equation~\ref{eq:emergence}. The hypothesis further asserts structure in the onset: collective capability appears sharply as interaction resources cross critical values, for example as shared bandwidth $\beta$ or restoration density $\nicefrac{1}{k}$ crosses a threshold predictable from the datasheet, so that emergence itself exhibits the phase-boundary phenomenology that cellular automata lead one to expect.
\end{description}

\subsection{Experimental program}

\paragraph{The frozen device.} \texttt{Cell v0.1} consists of a small frozen open language model (here, a $1.5$-billion-parameter model, with a $3$-billion-parameter model used as a second device \texttt{Cell v0.2}); scratchpad memory bounded at $H_{\max}$; a JSON-schema message interface with bandwidth $\beta$; a minimal perception--reason--act loop; and goals as declarative memory entries. The specification is published and versioned. Subsequent cell versions are new devices with new datasheets, never silent modifications.

\paragraph{Task families.} Four families instantiate the regimes. Decomposable--verifiable tasks, namely distributed sub-problems with mechanically checkable answers, instantiate the parallel regime. Sequential pipelines, namely multi-stage transformation and verification chains, instantiate the serial regime. Consensus and choice tasks, including naming games, instantiate the drift--selection regime. The fourth family instantiates the strong-emergence regime of Section~\ref{sec:emergence} and is constructed for $p \approx 0$ by design: distributed-evidence tasks whose answers require chaining $k$ evidence fragments spread across documents with $k \cdot |\text{fragment}| \gg H_{\max}$, so that the onset of collective capability can be traced as $k$ sweeps past the single-cell memory bound; long dependency chains with interleaved restoring stages, testing the predicted optimal spacing $k^{*}$ of Section~\ref{sec:laws}; and budgeted search tasks in which no single trajectory can reach a solution, run under three conditions, namely isolated cells with voting, cells sharing a search frontier, and cells exchanging intermediate lemmas, so that the differences across conditions measure $\Delta\Psi_{\mathrm{comm}}$ in pure form.

\paragraph{Signature experiments.} Two experiments anchor the program. The first is a full sweep over $N \in \{1, 2, 4, \dots, 1024\}$, over topologies including chain, star, random, and small-world graphs, and over bandwidth levels $\beta$ from low to high, measuring $\Psi$, $K$, $\rho$, $N_{\mathrm{eff}}$, $\Delta\Psi_{\mathrm{comm}}$, and consensus dynamics; its deliverables are the measured $\Psi$ surfaces, the fitted composition laws with datasheet-predicted against observed parameters, addressing H2 and H3, and the empirical drift--selection phase diagram. The second is the emergence-onset experiment on the fourth task family: with single-cell competence pinned at zero by construction, trace collective capability as the evidence span $k$, the bandwidth $\beta$, and the restoration density vary, producing the onset curves that test H7 and, if the onsets are sharp, the phase boundaries of collective cognition.

\paragraph{Baselines.} Three baselines are mandatory from the first experiment. A single cell with $N$-fold resampling and majority vote isolates the value of interaction via Equation~\ref{eq:deltapsi} and realizes the aggregation null. A single larger model at matched inference budget grounds H5. Published multi-agent frameworks at matched $N$ ground external validity.

\paragraph{Feasibility.} Every population experiment is inference-only over replicas of a small frozen model: embarrassingly parallel, requiring no model training, and executable on shared academic clusters. The collective-training line of Section~\ref{sec:learning} trains a single small model with parameter sharing, which recent low-cost reproductions of reasoning-oriented reinforcement learning have shown to be within modest academic budgets. This is by design. The framework's economics are aligned with the institutions most likely to pursue it.

\section{Illustrative Measurements}
\label{sec:results}

We instantiate the framework with a concrete frozen device and report a first round of measurements. The intent is not a benchmark, nor a complete execution of the agenda, but a demonstration that the datasheet and the null models are measurable and that they already discipline the questions the framework poses. Table~\ref{tab:claims} summarizes what these measurements support and what they leave open. Every result below is inference-only over replicas of a single frozen model.\footnote{An open-source runtime implements the cell, the population, the protocols, and the estimators. A deterministic backend that reproduces the datasheet parameters by construction is used only to validate the estimators and the pipeline; it is not evidence and is not reported here. All figures and numbers in this section are measured on frozen language models.}

\begin{table}[t]
\centering
\small
\caption{What this paper's measurements support, and what remains agenda. ``Consistent with'' denotes initial support on one cell family with modest compute, not confirmation.}
\label{tab:claims}
\begin{tabular}{@{}p{4.6cm}p{2.4cm}p{6.0cm}@{}}
\toprule
Claim & Status & Evidence / future work \\
\midrule
Datasheet is measurable; $\rho$ recoverable & consistent with (H2) & measured datasheets; estimator unbiased in calibration; full sufficiency test is future work \\
Voting helps only at low $\rho$; saturates near $N\!\approx\!1/\rho$ & consistent with (H1) & measured curves and $\rho$-intervention; single family, single topology so far \\
Binary voting model is a conservative lower bound & supported & measured vote exceeds the null as diversity grows; a dispersion-aware null is future work \\
Interactive protocols do not beat matched-cost voting & supported (this setting) & paired test on one family; broader protocols/tasks are future work \\
$\varphi(j)$ predicts the distributed-capacity ceiling & consistent with (H7) & two devices with an order-of-magnitude gap; realized onset not yet crossed \\
Topology sweep, consensus phase diagram, cost crossover (H5), collective training (H6) & not tested & specified in Section~\ref{sec:agenda}; future work \\
\bottomrule
\end{tabular}
\end{table}

\paragraph{Device, tasks, and estimator calibration.} \texttt{Cell v0.1} is the instruction-tuned Qwen2.5-1.5B model, frozen, with a $2048$-token scratchpad and a JSON message interface, served for batched inference. Where a more capable unit is needed for comparison, \texttt{Cell v0.2} is the frozen Qwen2.5-3B model under the same interface; it is a distinct device with its own datasheet, not a modification of v0.1. Tasks are generated procedurally so that answers are mechanically checkable and free of training contamination: compositional arithmetic, propositional-logic evaluation, and constrained counting for the aggregation regime, and a distributed-evidence task for the memory regime. Because the error correlation $\rho$ is the load-bearing datasheet parameter, we first verify that it is recoverable. On synthetic data with known $\rho$ drawn from a beta-binomial generator, the moment estimator we use is essentially unbiased at the operating point of our measurements ($M=16$ resamples over $I=60$ instances): the bias stays below $0.004$ across $\rho \in [0.05, 0.9]$, with root-mean-square error between $0.02$ and $0.06$. A measured $\rho$ is therefore trustworthy to within about $\pm 0.05$.

\paragraph{The datasheet of a cell.} Table~\ref{tab:datasheet} reports measured datasheets. Two qualitative facts stand out. First, a mandatory sanity check passes: at temperature zero, where every cell produces identical trajectories, the estimated error correlation is $\rho = 1.00$ for every family, confirming that $\rho$ measures the shared, systematic component of error rather than sampling noise. Second, the two devices differ in an instructive way. The larger cell (v0.2) is more competent on some families but markedly more \emph{peaked}: its output diversity $D$ falls to as little as $0.05$ bits and its error correlation rises to $0.85$--$0.94$, against $D \approx 1.3$ bits and $\rho \approx 0.6$ for the smaller cell. A more capable unit is not automatically a better \emph{population} member; higher confidence lowers diversity, raises $\rho$, and by Equation~\ref{eq:neff} saturates voting sooner. This capability--diversity tension is exactly the kind of design fact the datasheet is meant to expose.

\begin{table}[t]
\centering
\caption{Measured datasheets on procedurally generated task families (temperature $0.7$). Competence $p$ and correlation $\rho$ depend on task difficulty and are reported at the settings used below; $D$ is output entropy in bits, $\varphi$ single-fact communication fidelity, $\eta$ persuadability. Working-memory span $w$ did not collapse within the probed range ($\le H_{\max}$) for either cell. A temperature-zero control gives $\rho = 1.00$ in all rows.}
\label{tab:datasheet}
\begin{tabular}{llccccc}
\toprule
Cell & Family & $p$ & $D$ (bits) & $\rho$ & $\varphi$ & $\eta$ \\
\midrule
v0.1 (1.5B) & arithmetic & $0.37$ & $1.27$ & $0.58$ & $0.93$ & $0.07$ \\
v0.2 (3B)   & arithmetic & $0.43$ & $0.38$ & $0.92$ & $1.00$ & $0.25$ \\
v0.2 (3B)   & logic      & $0.58$ & $0.05$ & $0.94$ & $1.00$ & $0.25$ \\
v0.2 (3B)   & counting   & $0.76$ & $0.12$ & $0.85$ & $1.00$ & $0.25$ \\
\bottomrule
\end{tabular}
\end{table}

\begin{figure}[t]
\centering
\includegraphics[width=\textwidth]{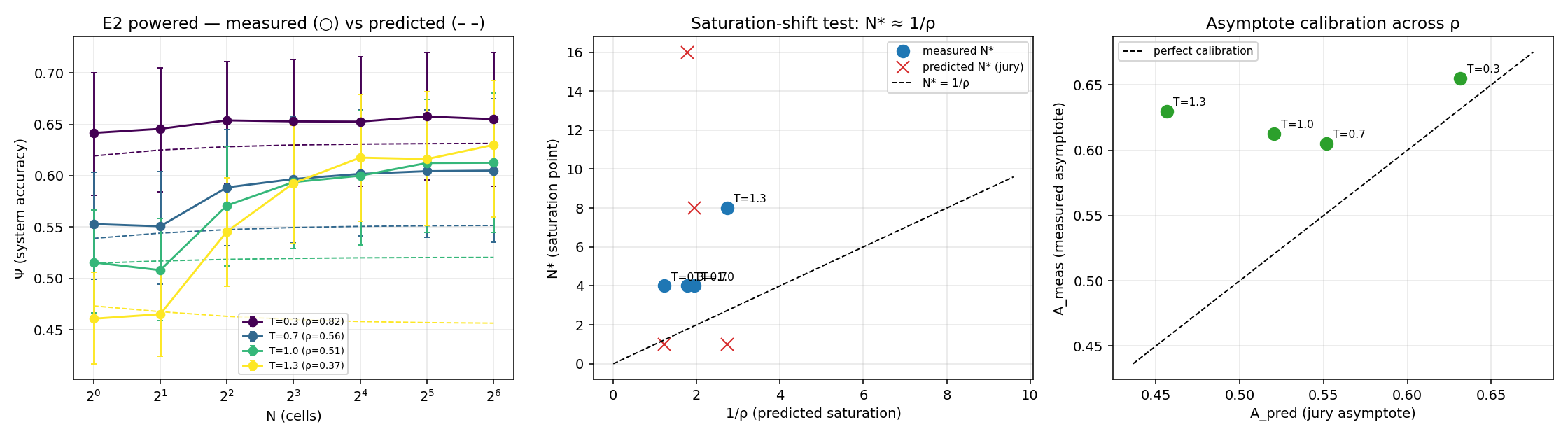}
\caption{Majority vote on frozen \texttt{Cell v0.1}. \emph{Left}: measured accuracy versus $N$ (solid, with $95\%$ bootstrap intervals) against the correct/incorrect voting prediction (dashed) at four temperatures; measured curves sit above the prediction and the gap grows as temperature raises diversity. \emph{Center}: the saturation point against $\nicefrac{1}{\rho}$. \emph{Right}: measured asymptotic accuracy against the predicted asymptote across temperatures; the measured value stays high while the binary prediction falls with competence, quantifying the conservatism of the null. All quantities are measured on the frozen model.}
\label{fig:voting}
\end{figure}

\paragraph{Voting and its null model.} We measure majority vote over $N \in \{1, 2, \dots, 64\}$ with no communication, on arithmetic instances at a difficulty giving single-cell competence $p \approx 0.65$ and correlation $\rho \approx 0.63$, and we intervene on $\rho$ by varying decoding temperature. Two findings emerge (Figure~\ref{fig:voting}). First, the qualitative predictions of the null model hold: gains are real but bounded, and lowering $\rho$ shifts the saturation point. As temperature rises from $0.3$ to $1.3$, the measured correlation falls from $\rho \approx 0.82$ to $\rho \approx 0.37$, and the accuracy--versus--$N$ curve saturates progressively later, as $N \approx \nicefrac{1}{\rho}$ predicts. Second, and more informatively, the correct/incorrect model is a \emph{conservative lower bound}: the measured asymptotic vote accuracy sits above the model's prediction, and the gap widens as diversity grows. At temperature $1.3$ the single-cell competence is $p = 0.47$, below the one-half at which the binary model predicts collective collapse, yet the vote reaches accuracy $0.63$, because the model's errors are dispersed across many wrong answers so the plurality of correct answers still wins. This is direct evidence for the scope caveat of Section~\ref{sec:capacity}: on open-ended tasks the aggregation null must account for error dispersion, and treated as a bound it is not violated but systematically exceeded.

\paragraph{The value of communication.} We next ask whether interaction beats aggregation at matched cost. Three popular protocols, round-based debate, a shared blackboard, and chain revision, are run so that their total inference cost in tokens matches that of the voting baseline, and each is compared to voting on the same instances. In our setting none beats voting (Table~\ref{tab:comm}). The paired difference between each protocol and the matched-cost vote is $-0.02$, with a $95\%$ interval of $[-0.10, +0.07]$ that includes zero; against the stronger null of spending the same token budget on additional independent votes, all three are worse by about $0.09$. The lesson is methodological rather than a dismissal of communication: reported against the right null, at matched cost, these protocols provide no benefit here, and a claim of benefit requires this comparison. A cell that genuinely exploited debate would show up as a positive value, which the framework makes the reported quantity.

\begin{table}[t]
\centering
\caption{Value of communication at matched token cost on \texttt{Cell v0.1} (arithmetic, $60$ instances). $\Psi$ is system accuracy; $\Delta\Psi_{\mathrm{comm}}$ is the change relative to spending the same token budget on independent votes; the last column is the paired difference from the matched-cost majority vote, with a $95\%$ bootstrap interval. No protocol beats voting.}
\label{tab:comm}
\begin{tabular}{lcccc}
\toprule
Protocol & Configuration & $\Psi$ & $\Delta\Psi_{\mathrm{comm}}$ & vs.\ vote [95\% CI] \\
\midrule
Majority vote (null) & $N=16$ & $0.52$ & --- & --- \\
Debate & $N=4$, $4$ rounds & $0.50$ & $-0.09$ & $-0.02$ $[-0.10, +0.07]$ \\
Blackboard & $N=4$, $4$ rounds & $0.50$ & $-0.09$ & $-0.02$ $[-0.10, +0.07]$ \\
Chain revision & $16$ stages & $0.50$ & $-0.09$ & $-0.02$ $[-0.10, +0.07]$ \\
\bottomrule
\end{tabular}
\end{table}

\begin{figure}[t]
\centering
\includegraphics[width=0.62\textwidth]{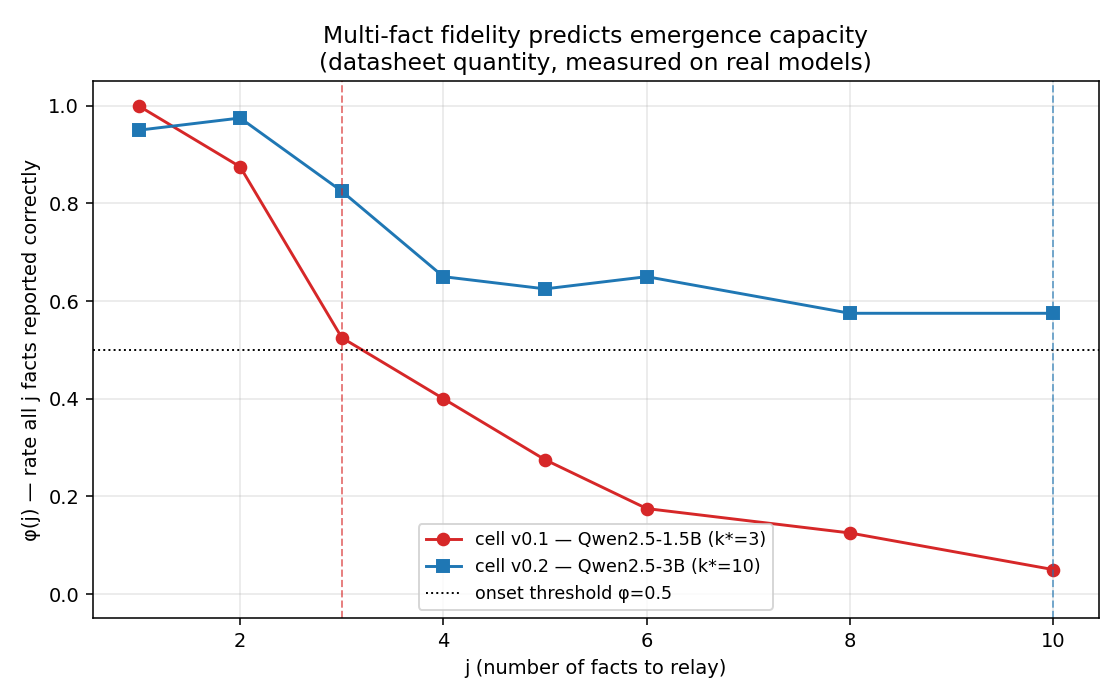}
\caption{Multi-fact fidelity $\varphi(j)$, the rate at which a cell correctly relays $j$ facts, measured on two frozen devices. \texttt{Cell v0.1} (1.5B) falls below one-half at $j \approx 3$; \texttt{Cell v0.2} (3B) stays above one-half through $j = 10$. This datasheet quantity predicts each cell's ceiling for tasks whose evidence exceeds a single cell's memory.}
\label{fig:phij}
\end{figure}

\paragraph{Distributed capacity and its measurable prerequisite.} The distributed-evidence task splits $k$ clue fragments across documents whose total size exceeds one cell's memory, so single-cell competence is null by construction and voting cannot help; whether a population solves it depends on cells relaying facts faithfully over rounds. We measure this prerequisite directly through a multi-fact extension of communication fidelity: $\varphi(j)$, the rate at which a cell correctly reports $j$ facts handed to it at once. For \texttt{Cell v0.1}, $\varphi(j)$ falls below one-half at $j \approx 3$; for the more capable \texttt{Cell v0.2} it remains above one-half through $j = 10$ (Figure~\ref{fig:phij}). The datasheet thus predicts a low ceiling on the distributed capacity of the small cell and a much higher one for the larger cell, and the small cell indeed fails the distributed task once $k$ exceeds about three, as its $\varphi(j)$ predicts.

We report an honest limitation at this frontier. Even the larger cell does not cross the onset cleanly end-to-end in our runs. Inspection shows that the cells extract a single injected fact reliably but degrade when accumulating several facts over rounds, because multi-round natural-language coordination is lossy: verbose intermediate messages accumulate and the running answer drifts. Realized distributed capacity therefore requires both high relay fidelity, which $\varphi(j)$ measures, and clean multi-round coordination, which small frozen cells lack. The mechanism and its prerequisite are measurable; crossing the onset robustly appears to require a more capable cell or a more constrained protocol. We regard this as a result about \emph{where} the onset sits, not as a negative one: it locates the missing ingredient precisely, in a datasheet term that separates cells by an order of magnitude in relay capacity.

\section{Discussion and Limitations}
\label{sec:discussion}

\paragraph{What the framework does not claim.} The vocabulary makes claims about architecture only, namely a repeated frozen unit whose replication supports scaling; it carries no mechanistic commitment about brains. The transistor and cellular-automata comparisons are motivations, not equivalences: a language model is a highly contextual, probabilistic device rather than a stable physical component, and a cognitive cell lacks the formal closure of a finite-state rule. Whether a compact datasheet nonetheless predicts its compositions is the empirical bet of H2. The datasheet-sufficiency claim of H2 is a bet, and the framework is built so that its failure is informative: the dimensions along which a compact datasheet cannot predict collective behavior would mark where cognitive units differ in kind from electronic ones. The fixed-cell principle trades breadth for compositional validity. Systems built under it will underperform per-task-engineered agents on any single benchmark, and the framework declines that comparison, as device physics declines to compete with hand-tuned circuits.

\paragraph{Known pathologies as objects of study.} The literature already documents the failure modes a device theory must predict: effective-population collapse under error correlation, which is the machine analog of the Ringelmann effect; non-monotonic returns to additional calls \cite{chen2024more}; cascade error amplification in pipelines; superlinear coordination overhead; and drift-dominated consensus \cite{tanaka2026collective}. Each appears in this framework as a measurable regime with named control parameters, and each is, on its own, a publishable study of mechanism and mitigation.

\paragraph{Relation to adjacent programs.} The framework complements research on test-time compute, since a population is a structured inference-time compute allocation \cite{snell2024scaling}, and on agent memory, which supplies the candidate mechanism of specialization in H4. Its nearest relatives are the cellular-automata methodology of Section~\ref{sec:ca}, whose stance it imports; the minimal-model program of \cite{tanaka2026collective}, whose statistical-mechanics style it adopts for the consensual regime; and the collaborative-scaling line of \cite{qian2024scaling}, whose law it seeks to derive from device parameters.

\paragraph{Threats to validity.} Three deserve statement. First, datasheet parameters may be task-family-dependent to a degree that erodes their predictive economy. The program mitigates this by reporting $d(\mathcal{C}; \mathcal{T})$ per family and studying the stability of the map across families. Second, results obtained with small cells may fail to extrapolate to more capable units. The versioned-device discipline turns this into a measurable question, namely how composition-law parameters move across cell versions. Third, the null model of Equation~\ref{eq:jury} presumes exchangeability, which structured topologies break. The appropriate generalizations, correlated juries on graphs, are known territory in collective decision theory and constitute theory work the agenda explicitly includes.

\section{Conclusion}
\label{sec:conclusion}

This paper proposed the cognitive cell, a minimal, frozen, versioned unit of cognition with explicit interfaces for perception, memory, reasoning, action, goals, and communication, as a device abstraction for compositional artificial cognition, and organized around it a research program with three pillars: characterization through the datasheet; candidate composition laws for the parallel, serial, and consensual regimes; and the mapping of tasks and budgets to populations. The framework carries the methodology of cellular automata to a more expressive unit and subsumes multi-agent architectures as the special case of autonomous cells. An initial round of measurements on a small frozen model supports the framing: the datasheet is measurable; adding cells helps only when their error correlation is low, so that an apparently contradictory literature becomes consistent under a single measured parameter; the simplest correct/incorrect voting model is a conservative lower bound that open-ended voting exceeds through error dispersion; popular interactive protocols do not beat matched-cost voting in our setting; and a datasheet quantity, multi-fact fidelity, predicts the prerequisite for capability that no single cell has, while locating precisely why small frozen cells do not yet realize it. The hypotheses are falsifiable, the experiments are inference-parallel and affordable, and the intended products, namely datasheets, laws, and an open runtime, are the kind that compound across studies and across groups. The proposal, in the end, is a change of question: what is the smallest unit that composes, and what laws govern the composition.

\section*{Disclosure of AI Assistance}
The author used Claude Fable 5 (Anthropic) to assist with literature survey, drafting, and \LaTeX{} preparation. The core concept, research direction, and all content decisions were the author's, who reviewed and verified all claims, derivations, and references, and takes sole responsibility for the final content.

\bibliographystyle{unsrtnat}
\bibliography{references}

\end{document}